\documentclass[prl,aps,showpacs,epsf, twocolumn]{revtex4}

\usepackage[dvips]{graphicx}

\def\be{\begin{equation}}
\def\ee{\end{equation}}

\usepackage{bm}

\begin{document}

\title{Low lying twisting and acoustic modes of a rotating Bose-Einstein condensate}
\author{F. Chevy}
 \affiliation{Laboratoire Kastler-Brossel$^*$,
ENS, 24 rue Lhomond, 75005 Paris}

\date{\today}

\begin{abstract}
We present a calculation of the low lying spectrum of a rotating
Bose-Einstein condensate. We show that in a cylindrical geometry,
there exist two linear branches, one associated with usual
acoustic excitations, the other corresponding to a twisting mode
of the vortex lattice. Using a hydrodynamical approach we derive
the elasticity coefficient of the vortex lattice and calculate the
spectrum of condensate in a three dimensional harmonic trap with
cylindrical symmetry.
\end{abstract}

 \maketitle

By contrast with classical hydrodynamics, a quantum fluid cannot
rotate like a solid body. Instead, it carries quantized vortices
self organizing along a triangular Abrikosov lattice when the
rotation is fast enough. Vortices constitute a universal
characteristic of quantum fluids and were directly observed in
systems as different as liquid helium \cite{Donnelly91},
superconductors \cite{Cribier64}, gaseous Bose-Einstein
condensates (BEC) \cite{Cornell00,Madison00,AboShaeer01,Hodby02}
and very recently fermionic superfluids \cite{Zwierlein05}. The
study of the excitations of a vortex lattice was  initiated by the
work of Tkachenko \cite{Tkatchenko66} that was restricted to the
study of the propagation of waves transversely to the rotation
axis in an incompressible superfluid. The full understanding of
these excitations in the case of dilute gases remains a challenge,
due to the non trivial interplay between the elasticity of the
vortex lattice and the phonon modes associated with the
compressibility of these systems. Partial results were obtained in
the case of a single vortex line \cite{Pitaevski61,Martikainen03},
two-dimensional systems \cite{Anglin,Mizushima04,Baksmaty03}, fast
rotating systems \cite{Baym03,Cozzini04,Gifford04} or using the
rotational hydrodynamics formalism
\cite{Sedrakian01,Chevy03,Choi03}, but a complete theory remains
to be found.

\begin{figure}
\includegraphics[width=\columnwidth]{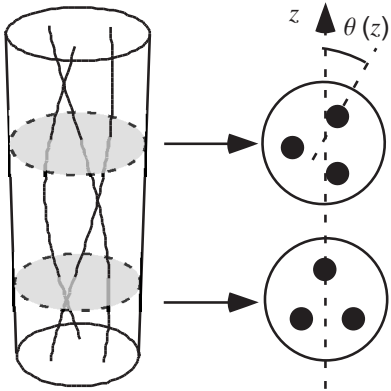} \caption{Left: Structure of the twiston mode for a cylindrical
BEC carrying three vortex lines. Right: arrangement of the vortex
lattice in the planes defined by the shaded areas. The vortex
lattice is rotated by an angle $\theta(z)$ depending on the axial
position along the trap. This twisting induces the propagation of
an elastic wave along the condensate axis.} \label{Fig2}
\end{figure}

In this letter, we present a study of the low lying modes
propagating along the axis of an elongated vortex lattice. Up to
now, these modes have only been considered for homogeneous and
unbounded fluids \cite{Gifford04}. In the more realistic case of a
trapped gaseous BEC, we show that these modes can be understood as
Goldstone modes arising from  U(1) and O(2) broken symmetries,
 associated respectively with the choice of the phase
of the macroscopic wave function of the condensate and of the
direction of the vortex lattice. These two broken symmetries thus
give rise to two low energy excitation branches associated
respectively to a modulation of the phase of the wave function and
of the direction of the vortex lattice in the $z$ direction. Since
the gradient of the phase is proportional to the velocity of the
BEC, the first branch is associated with an acoustic wave
propagating along the cloud (``phonon" branch). The second mode
corresponds to the twisting of the vortices and is related to the
elasticity of the lattice (``twiston" branch, Fig. \ref{Fig2}).
This scheme will be worked out first using a perturbative
resolution of the Bogoliubov-de Gennes equations in a cylindrical
trap, a method already used in the stability analysis of solitons
\cite{Kuznetsov95}. We will show that this solution can be
interpreted in a macroscopic framework, leading to
elasto-hydrodynamical equations for the motion of the BEC in the
presence of the vortex lattice. In particular, we propose the
first derivation of the elastic response coefficient of the vortex
lattice, starting from first principles. Finally, local density
approximation will allow for the calculation of the lowest energy
modes  in the presence of an axial trapping.

We consider a dilute Bose-Einstein condensate rotating at an
angular velocity $\Omega_0$ along the $z$ axis. At rest, the
system is described  in the mean-field approximation by a
macroscopic wave-function $\psi_0$ solution of the
Gross-Pitaevskii equation

\be \left(\widehat h_0+g|\psi_0|^2\right)\psi_0=0. \label{Eqn1}\ee

\noindent where the single particle hamiltonian $\widehat h_0$ is
given by $\widehat h_0=-\frac{\hbar^2}{2m}\nabla^2+V(x,y)-\Omega_0
\widehat L_z-\mu_0$. $V$ is the transverse trapping potential,
$\widehat L_z$  the angular momentum in the $z$ direction, $\mu_0$
the chemical potential and $g$ the coupling constant
characterizing the 2-body interactions. We assume for the moment
there is no confining potential in the $z$ direction.
Nevertheless, we impose periodic boundary conditions in this
direction,  with period $L$, to allow for the normalisation of the
wave-function.

In the linear regime, we write function $\psi$ of the condensate
is given by $\psi=\psi_0+\delta\psi$. $\delta\psi$ is then
solution of the Bogoliubov de-Gennes equations
$i\hbar\partial_t\bm\Phi=\widehat{\cal L}\bm\Phi$, with
$\bm\Phi=(\delta\psi,\delta\psi^*)$ and


\be\widehat{\cal L}= \left(
\begin{array}{cc}
\widehat h_0+2g|\psi_0|^2&g\psi_0^2\\
-g{\psi_0^*}^2&-\widehat h_0-2g|\psi_0|^2
\end{array}
\right).\ee

We recall a few properties  of $\widehat{\cal L}$ relevant for the
discussion \cite{Lewenstein96,Castin98}. First, although
$\widehat{\cal L}$ is not hermitian, it is orthogonal for the
quadratic form with signature (1,-1) defined by
$(\bm\Phi_1|\bm\Phi_2)=\int d^3\bm r
\bm\Phi_1^\dagger\widehat\sigma_3\bm\Phi_2$, where
$\widehat\sigma_3$ is the diagonal Pauli matrix. Second, the
eigenvectors $\bm\Phi_\alpha$ of $\widehat{\cal L}$ do not
constitute a complete basis. Indeed, one can show that each zero
energy mode is associated with an anomalous Jordan mode
$\bm\Phi'_\alpha$ such that $\widehat{\cal
L}\bm\Phi'_\alpha=\bm\Phi_\alpha$. In the case under study here,
we assume there are only two zero energy modes associated
respectively with the U(1) and O(2) freedom of choice of the phase
of the wave function and the orientation of the vortex lattice.
The associated eigenvectors are respectively denoted by
$\bm\Phi_p$ (phase symmetry) and $\bm\Phi_r$ (rotational symmetry)
and are given by \cite{Lewenstein96,Castin98}

\be
\begin{array}{ll} \bm\Phi_p=\left(
\begin{array}{c}
\psi_0\\
 -\psi_0^*
\end{array}
\right) & \bm\Phi'_p=\left(
\begin{array}{c}
\partial_\mu\psi_0\\
\partial_\mu\psi_0^*
\end{array}
\right)\\
\bm\Phi_r=\left(
\begin{array}{c}
\widehat L_z\psi_0\\
-(\widehat L_z\psi_0)^*
\end{array}
\right)&\bm\Phi'_r=\left(
\begin{array}{c}
\partial_\Omega\psi_0\\
 \partial_\Omega\psi_0^*
\end{array}
\right),
\end{array}
\label{Eqn16} \ee

 Let us now proceed with the calculation of the long wavelength
modes of $\widehat{\cal L}$. Using the translational invariance of
the system we can write $\bm \Phi_\alpha (x,y,z) =\bar{\bm
\Phi}_\alpha (x,y) e^{ik_\alpha z}$. We therefore see that
$\widehat{\cal L}=\widehat{\cal L}_0+\delta\widehat{\cal L}$,
where $\widehat{\cal L}_0$ acts on the transverse degrees of
freedom only, $\delta \widehat{\cal
L}=\epsilon_{k_\alpha}\widehat\sigma_3$, with
$\epsilon_{k_\alpha}=\hbar^2k_\alpha^2/2m$.

Since we only care for  long wavelength eigenmodes, we have
$\epsilon_{k_\alpha}$ vanishingly small and we can treat $\delta
\widehat {\cal L}$ as a perturbation of $\widehat {\cal L}_0$.
However, starting the perturbation expansion from the zero energy
modes $\bm\Phi_r$ and $\bm\Phi_p$ of ${\cal L}_0$ associated with
the rotational and phase symmetries gives rise to {\em linear}
excited branches. Indeed, due to the non diagonalizability  of
$\widehat {\cal L}_0$ the first order term of the low momentum
expansion scales like $\epsilon_{k_\alpha}^{1/2}$
 rather than $\epsilon_{k_\alpha}$ \cite{Lobo05}. This singularity can be proven
  rigorously by diagonalizing the projection of $\delta\widehat {\cal L}$ on the vector space
 spanned by the zero energy and anomalous modes. Physically,
 this result corresponds to the fact that one should recover a
linear phonon dispersion law $E_\alpha\propto k_\alpha\propto
 \epsilon_{k_\alpha}^{1/2}$ associated with usual acoustic waves.

Working out the perturbative expansion in power of
$\sqrt{\epsilon_{k_\alpha}}$ yields

\be \bar{\bm\Phi}_\alpha
=(a_r\bm\Phi_r+a_p\bm\Phi_p)+E_\alpha(a_r\bm\Phi'_r+a_p\bm\Phi'_p)+...,
\label{Eqn18}\ee

\noindent where $E_\alpha$ is solution of the eigenequation

\be \sum_{\gamma =r,p}(\epsilon_{k_\alpha}
A_{\beta\gamma}-E_\alpha^2B_{\beta\gamma})a_\gamma=0,\label{Eqn5}\ee

\noindent with $\beta\in\{r,p\}$,
$A_{\beta\gamma}=(\bm\Phi_\beta|\widehat\sigma_z\bm\Phi_\gamma)/L$,
and $B_{\beta\gamma}=(\bm\Phi_\beta|\bm\Phi'_\gamma)/L$. $L$ is
introduced here so that $A$ and $B$ are defined as 1D linear
quantities that we can use in a local density approach as
demonstrated later in this paper. Using Eq. \ref{Eqn16} yields the
simple expressions

\be\begin{array}{ll} (\bm\Phi_p|\widehat\sigma_z\bm\Phi_p)=2N&
(\bm\Phi_p|\bm\Phi'_p)=\partial_{\mu_0} N\\
(\bm\Phi_r|\widehat\sigma_z\bm\Phi_r)=2\langle\widehat
L_z^2\rangle& (\bm\Phi_r|\bm\Phi'_r)=\partial_{\Omega_0}
\langle\widehat
L_z\rangle\\
(\bm\Phi_p|\widehat\sigma_z\bm\Phi_r)=2\langle\widehat L_z\rangle&
(\bm\Phi_p|\bm\Phi'_r)=\partial_{\Omega_0} N\\
(\bm\Phi_r|\widehat\sigma_z\bm\Phi_p)=2\langle\widehat
L_z\rangle&(\bm\Phi_r|\bm\Phi'_p)=\partial_{\mu_0} \langle\widehat
L_z\rangle,\\
\end{array}
\label{Eqn14} \ee

\noindent where $N$ is the total atom number \cite{Symmetry}. The
set of two equations (\ref{Eqn5}) yields two different linear
excitation branches. The highest branch has a non-zero velocity
for vanishing $\Omega_0$ and can therefore be identified with a
phonon mode. The other branch has a vanishing velocity at small
$\Omega$ and is therefore the twiston mode.

It is striking that the coefficients of Eq. (\ref{Eqn5}) can be
expressed as simple combinations of macroscopic quantities such as
the atom number or the angular momentum. Similarly to superfluid
hydrodynamics, this suggests that the formalism developed here can
be expressed in term of an elasto-hydrodynamical theory. In order
to clarify this link, we first note that the frequency of the long
wavelength modes we are interested in is much smaller than the
frequencies characterizing the evolution of the transverse degrees
of freedom. This means in particular that the dynamics in the
$(x,y)$ plane is frozen and that we can therefore define local
chemical potential $\mu(z,t)$, angular velocity $\Omega (z,t)$,
phase of the wavefunction $\chi(z,t)$ and angle of the vortex
lattice $\theta (z,t)$. In other word, the wave function can be
written as

\begin{equation}
\psi(x,y,z,t)=\psi_0(\mu(z,t),\Omega(z,t),\chi(z,t),\theta(z,t),x,y)
\end{equation}

\noindent where $\psi_0$ is the solution of the Gross Pitaevskii
equation (\ref{Eqn1}). Expanding the macroscopic quantities around
their equilibrium values yields

\begin{equation}
\delta\psi=\delta\mu\partial_{\mu_0}\psi_0+\delta\Omega\partial_{\Omega_0}\psi_0+\delta\chi\partial_{\chi_0}\psi_0+\delta\theta\partial_{\theta_0}\psi_0,
\label{Eqn8}
\end{equation}

\noindent where $\mu=\mu_0+\delta\mu$,
$\Omega=\Omega_0+\delta\mu$, etc. Noting further that $\delta\psi$
is the first component of $\bm\Phi_\alpha$ and that
$\partial_\chi\psi_0=i\psi_0$ and
$\partial_\theta\psi_0=-i\widehat L_z\psi_0/\hbar$, we see that
Eqn. (\ref{Eqn8}) is equivalent to

\be
\bm\Phi=i\delta\chi\bm\Phi_p-i\frac{\delta\theta}{\hbar}\bm\Phi_r+\delta\mu\bm\Phi'_p+\delta\Omega\bm\Phi'_r.
\ee

Comparing to Eq. (\ref{Eqn18}) this leads to the identification

\begin{eqnarray}
a_r=-i\delta\theta/\hbar=\delta\Omega/E_\alpha\\
a_p=i\delta\chi=\delta\mu/E_\alpha,
\end{eqnarray}

\noindent     which gives, using the identity
$\partial_t=-iE_\alpha/\hbar$,

\begin{eqnarray}
\partial_t\delta\theta&=&\delta\Omega\label{Eqn21}\\
\hbar\partial_t\delta\chi&=&-\delta\mu\label{Eqn22}
\end{eqnarray}

The interpretation of these two relations is straightforward.
Indeed, eq. \ref{Eqn21} is the analogue of the Kelvin-Helmholtz
theorem \cite{KH} and implies that the vortex lattice rotates at
the same speed as the local velocity flow. As for eq. \ref{Eqn22},
it is a version of the Bernoulli theorem, with $\hbar\delta\chi/m$
playing the role of the velocity potential.

Starting from the identification of the $a_{r,p}$ coefficient, we
now show that the eigensystem (\ref{Eqn5}) can be interpreted as
conservation laws for the system. Indeed, the conservation of
particle number, momentum  and angular momentum in the $z$
direction yield at first order in perturbation the very general
set of equations

\begin{eqnarray}
\partial_t\delta n+\partial_z (n_0 v)=0 \label{Eqn10}\\
mn_0\partial_t v+\partial_z f=0\label{Eqn11}\\
\partial_t\delta \ell_z+\partial_z\Gamma=0\label{Eqn12}
 \end{eqnarray}

\noindent where $n_0$ is the linear particle density, $v$ is the
local velocity in the $z$ direction, $\ell_z$ is the angular
momentum per unit length. $f$ and $\Gamma$ are respectively the 1D
momentum and angular momentum currents, which can be identified
with a force and a torque and will be calculated from the
microscopic formalism presented above.

Combining Eq. (\ref{Eqn10}) and (\ref{Eqn11}) and using a
generalized Gibbs-Duhem relation \cite{Kittel} for a rotating
system $df=n_0 d\mu+\ell_z^0 d\Omega$ yields

\be
\partial^2_t\left[\delta\mu\partial_{\mu_0}
n_0+\delta\Omega\partial_{\Omega_0}
n_0\right]=\frac{1}{m}\partial_z\left[n_0\partial_z\delta\mu+\ell_z^0\partial_z\delta\Omega\right]
\label{Eqn13}\ee

\noindent where we have used the adiabatic following of the
transverse degrees of freedom to set $\delta
n=\delta\mu\partial_{\mu_0} n_0+\delta\Omega\partial_{\Omega_0}
n_0$.

After making the replacement $\hbar\partial_t=-iE_\alpha$ and
$\partial_z=ik_\alpha$, we see that Eq. (\ref{Eqn13}) is strictly
equivalent to the $\alpha=p$ component of Eq. (\ref{Eqn5}).

A similar analysis shows that the $\alpha=r$ component of Eq.
(\ref{Eqn5}) is equivalent to the angular momentum conservation if
the current $\Gamma$ is taken to be

\be \Gamma=\ell_z^0 v-\kappa\partial_z\delta\theta,
\label{Eqn15}\ee

\noindent where $\kappa=\Delta L_z^2/Lm$, and $\Delta
L_z^2=\langle \widehat L_z^2\rangle-\langle \widehat
L_z\rangle^2/N$ is the fluctuation of the angular momentum. The
term of $\Gamma$ proportional to the local velocity $v$
corresponds the convective part of the angular momentum current.
The second term is associated with the elastic response of the
vortex lattice to a torsion and $\kappa$ is therefore the elastic
modulus of the lattice.

The hydrodynamical formalism developed above can be used to extend
our calculation to the case of a weak trapping in the $z$
direction. In this case, local density approximation can be
applied. At equilibrium, the angular velocity is still uniform
along the cloud. By contrast, the local chemical potential is now
given by $\mu_0 (z)=\mu_c-m\omega_z^2 z^2/2$, where $\mu_c$ is the
chemical potential at the center of the trap. This permits to
define local matrices $A(z)$ and $B(z)$ and yields the equation
for the vector $\bm X(z,t)=(\delta \mu(z,t),\delta\Omega(z,t))$

\be 2m\partial_t^2\left[B(z)\cdot\bm
X(z,t)\right]=\partial_z\left[A(z)\cdot\partial_z\bm X(z,t)\right]
\label{Eqn20}\ee

\begin{figure}
\includegraphics[width=\columnwidth]{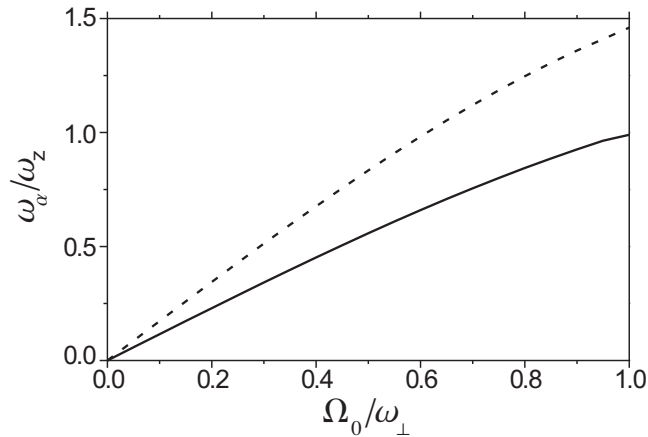}
\caption{Low lying twiston spectrum of a trapped rotating
Bose-Einstein condensate. The two branches correspond to the
frequency of the lowest odd (full line) and even (dashed line)
twiston modes and were calculated using polynomials trial wave
functions.} \label{Fig1}
\end{figure}

In general, $A$ and $B$ must be calculated using a numerical
resolution of the Gross-Pitaevskii equation. However, in the
regime of fast rotation, quantities appearing in Eq. (\ref{Eqn20})
can be evaluated assuming a classical rotationnal flow $\bm
v=\bm\Omega_0\times\bm r$. We have first checked that at zero
angular velocity we could recover the spectrum of an elongated
condensate studied in \cite{Stringari98}. We have also verified
that the center of mass was indeed evolving at a frequency
$\omega_z$, in accordance with Kohn's theorem. The other
eigenfrequencies $\omega_\alpha$ are evaluated using a variational
method. We indeed note that the $\omega_\alpha^2$ are the
eigenvalues of the operator $\widehat
T=B^{-1}\cdot\partial_z(A\cdot\partial_z\cdot)/2m$ which is
hermitian for the scalar product defined by $\langle\bm Y|\bm
X\rangle=\int \bm Y^\dagger(z)\cdot B(z)\cdot \bm X(z)\,dz$. We
have calculated the $\omega_\alpha$ using polynomial trial wave
functions of order 5. The variation of $\omega_\alpha$ for the two
lowest twiston modes -- corresponding respectively to odd and even
symmetries \cite{Projection} -- is displayed in Fig. \ref{Fig1}.

In conclusion, we have demonstrated the existence of two low
energy excitation branches of a rotating elongated Bose-Einstein
condensate. The validity of our calculation is limited by two
conditions. First the applicability of the hydrodynamical
formalism to evaluate the matrix elements of $A$ and $B$ requires
a dense vortex lattice, hence a large rotation frequency. Using an
imaginary time evolution of  2D Gross-Pitaevskii equation
\cite{Castin99}, we have checked it was true for $\Omega_0\gtrsim
0.8\omega_\perp$. Second the breakdown of the perturbation
expansion will happen when the $\omega_\alpha$ become comparable
with the frequency $\omega_T$ of the first transverse excited
mode, {\em i.e.} the lowest Tkachenko mode. Using the expression
of $\omega_T$ found in \cite{Cozzini04}, we find that for typical
values $\mu/\hbar\omega_\perp=20$ and $\omega_\perp/\omega_z=20$,
the perturbation expansion fails for
$\Omega_0/\omega_\perp\lesssim 0.96$. Although the width of
validity of our approximation might seem narrow at first sight, it
must be noted that experimentally regular vortex lattices are only
observed on a relatively narrow range of rotation frequencies,
comparable to our validity domain ($\Omega_0/\omega_\perp\in
[0.7,0.96]$ for \cite{Bretin04}). A last issue concerns the
experimental observation of twistons. It must be noted that, in
practice, it is difficult to excite directly the vortex degrees of
freedom of a gaseous BEC and it is therefore more convenient to
excite an acoustic mode by perturbing the trapping potential and
to use the linear \cite{Coddington03,Smith04} or non linear
\cite{Bretin03} coupling to vortical modes. In principle, the
coupling existing between acoustic and torsional degrees of
freedom should permit an excitation of the twiston mode using an
external potential. However, one can show using perturbation
theory that a simple modulation of the trap center or frequency is
only weakly coupled to the twiston mode. To obtain a significant
coupling to torsional modes, one therefore needs to impart cubic
or quartic perturbation to the gas.

{\acknowledgments The author specially thanks S. Stringari for
initiating this work and acknowledges fruitful discussion with G.
Baym, Y. Castin, and H.T.C. Stoof as well as the Cold atom group
of the ENS. He is also very grateful to J. Dalibard, D.
Gu\'ery-Odelin and C. Salomon for careful reading and suggestions.
This work is partially supported by CNRS, Coll\`{e}ge de France,
ACI Nanosciences and R\'{e}gion Ile de France (IFRAF).}


\begin{thebibliography}{99}
\bibitem{Symmetry} Note that $B_{\alpha\beta}$ is symmetric since
 Hellman-Feynman theorem yields $\partial_{\Omega_0}
N=\partial_{\mu_0}\langle\widehat
L_z\rangle=-\partial^2_{\Omega_0\mu_0}\langle \widehat H\rangle$.
\bibitem{Projection} Note that the real lowest even mode has zero
frequency and corresponds to $\delta\mu$ and $\delta\Omega$
constant. To get rid of it, we have carried out the variationnal
method in the vector space orthogonal to constant polynomials.
\end{thebibliography}
\end{document}